# Delocalization in infinite disordered 2D lattices of different geometry

**E G Kostadinova, K Busse, N Ellis, J Padgett, C D Liaw, L S Matthews, and T W Hyde**
Baylor University, 101 Bagby Avenue, Waco, TX, 76706, USA

E-mail: Eva_Kostadinova@baylor.edu, Kyle_Busse@baylor.edu, Naoki_Ellis@baylor.edu, Josh_Padgett@baylor.edu, Constanze_Liaw@baylor.edu, Lorin_Matthews@baylor.edu, Truell_Hyde@baylor.edu

**Abstract.** The spectral approach to infinite disordered crystals is applied to an Anderson-type Hamiltonian to demonstrate the existence of extended states for nonzero disorder in 2D lattices of different geometries. The numerical simulations shown prove that extended states exist for disordered honeycomb, triangular, and square crystals. This observation stands in contrast to the predictions of scaling theory, and aligns with experiments in photonic lattices and electron systems. The method used is the only theoretical approach aimed at showing delocalization. A comparison of the results for the three geometries indicates that the triangular and honeycomb lattices experience transition in the transport behavior for same amount of disorder, which is to be expected from planar duality. This provides justification for the use of artificially-prepared triangular lattices as analogues for honeycomb materials, such as graphene. The analysis also shows that the transition in the honeycomb case happens more abruptly as compared to the other two geometries, which can be attributed to the number of nearest neighbors. We outline the advantages of the spectral approach as a viable alternative to scaling theory and discuss its applicability to transport problems in both quantum and classical 2D systems.

# I. INTRODUCTION

Due to its wide applicability, the subject of Anderson localization has grown into a rich field across both physics and mathematics. In condensed matter, a crystal without impurities at zero temperature acts like a perfect conductor for a travelling wave-particle. According to Anderson [1], as the level of impurities reaches a critical value, the conductance of the crystal decreases and the travelling wave-particle experiences a transition from an extended to an exponentially localized state, called Anderson localization. Anderson localization is currently well understood for the 1D [2]–[4] and 3D [5]–[10] cases, where the problem has been studied in both quantum mechanical [5], [8], [11]–[14] and classical [15]–[21] systems. However, the existence of metal-to-insulator transition (MIT) in the critical 2D case has been the subject of heated debate over the past few decades due to a disagreement between theoretical prediction and experimental observation. According to the well-established scaling theory [22], all electron states in a 2D crystal should be exponentially localized for any amount of disorder. Nevertheless, a transition from strong to weak localization has been observed in photonic lattices [23], [24] and a MIT has been reported in electron systems [25]–[28].

With the discovery of graphene in 2004 [29], [30], the resolution of the 2D transport debate has become increasingly important. The extraordinary properties of graphene make it an ideal candidate for the development of flexible and durable semi-transparent technology, improved energy storage units, high-efficiency solar panels, and water purification systems. The realization of such applications is sensitive to a deep understanding of the transport properties of this 2D material. Recently, a metal-to-insulator transition has been experimentally established for graphene doped with hydrogen [31]. However, the origins of this transition are unclear. One possible explanation is that graphene exhibits a localization length longer than the size of current numerical simulations, which leads to the "apparent" existence of delocalized states [32]–[34]. The other major hypothesis, which will be the one examined in this work, is that scaling theory does not yield reliable results for all dimensions and system sizes [35], [36], and that its limitations come to play in the critical 2D case.

In physical applications, there are two widely used numerical methods for studying Anderson localization: the Kubo-Greenwood (KG) theory [37], [38] and the recursive Green's function technique (RGT) [39], [40]. Both approaches rely on finite size scaling and periodic boundary conditions to restrict the energy of a random Hamiltonian acting on an unbounded lattice. Since the spectra (and therefore important physical properties) of self-adjoint operators are sensitive to boundary conditions applied to the edges of the domain, restriction of the Hilbert space can lead to false information about the energy states of the Hamiltonian. For example, one unfortunate side effect of any finite size formulation of the Hamiltonian is that the existence of scattering states (in the sense of absolutely continuous spectrum) is excluded a priori.

In this paper, we apply the (recently introduced) spectral approach [41]–[43] to study delocalization in infinite disordered 2D systems and demonstrate the existence of extended states (and, therefore, MIT) in 2D honeycomb, triangular, and square lattices. The spectral approach employs a bounded Hamiltonian, which is defined on the entire Hilbert space without requiring scaling or the assumption of periodic boundary conditions. The use of spectral theory has been recognized by mathematicians as a valid approach to the Anderson localization problem [44]–[46] but has yet to be applied to physical systems. Here we argue that the results of the spectral approach can significantly contribute to the debate over the transport properties of 2D materials.

This paper is arranged in the following manner. We start with an introduction of the Anderson localization problem and the spectral approach to its solution (Sec. II). We then present results from applying the spectral approach to numerical simulations of the 2D honeycomb, triangular, and square lattices (Sec. III). In this section, we improve the analysis method used in our previous work [43] and use it to identify the critical amount of disorder that marks the onset of a metal-to-insulator transition for each geometry. A comparison between the transport properties of the honeycomb and the triangular lattices is presented in Sec. IV. Finally, we discuss the application of the spectral approach to the quantum percolation problem and to the classical 2D complex plasma crystal (Sec. V).

## II. THEORETICAL BACKGROUND

### A. Formulation of the Anderson problem

Here we introduce the Anderson localization problem using the square lattice as an example. The same analysis can easily be generalized to any lattice geometry by an appropriate choice of unit vectors of the corresponding Hilbert space. Consider the separable Hilbert space $l^2(\mathbb{Z}^2)$ of square-summable sequences on the square lattice $\mathbb{Z}^2$ (see Fig. 1c). The discrete random Schrödinger operator in 2D is given by the self-adjoint Hamiltonian operator

$$H_\varepsilon = -\Delta + \sum_{i \in \mathbb{Z}^2} \varepsilon_i \delta_i \langle \delta_i |, \tag{1}$$

where $\Delta$ is the discrete Laplacian[1] in 2D, $\delta_i$ are the standard basis vectors of $\mathbb{Z}^2$ ($\delta_i$ assumes the value 1 in the $i^{th}$ entry and zero in all other entries), and $\varepsilon = \{\varepsilon_i\}_{i \in \mathbb{Z}^2}$ is a set of random variables on a probability space $(\Omega, P)$. The random variables $\varepsilon_i$ are independent identically distributed (i.i.d.) in the interval $[-W/2, W/2]$ with probability density $\chi$, satisfying

$$\chi(\tau) = \begin{cases} 0, & \tau \notin [-W/2, W/2] \\ 1/W, & \tau \in [-W/2, W/2] \end{cases}. \tag{2}$$

In other words, the random variables are uniformly distributed, so that $\varepsilon_i$ takes each value in $[-W/2, W/2]$ with equal probability. In particular, here we have

$$\int_0^{W/2} \chi(\varepsilon) d\varepsilon = \frac{1}{2} \tag{3}$$

The Hamiltonian in (1) models an infinite 2D lattice comprised of atoms located at the integer points $\mathbb{Z}^2$. Each lattice site is assigned a random amount of energy $\varepsilon_i$, which defines the disorder in the system. The magnitude of the disorder is determined by the width of the interval $[-W/2, W/2]$, i.e. by the value of $W$.

For a given value of the disorder $W$, one is interested if the energy spectrum of the given Hamiltonian consists of localized or extended states. In our previous work [43] we showed that in the case of a 2D square lattice, extended states occur for $W \leq 0.60$. In this paper, we improve the accuracy of this result and further show delocalization for both honeycomb and triangular lattices, thus confirming the existence of extended states for various 2D geometries.

---

[1] *The discrete Laplacian is the analogue of the continuous Laplacian on a graph or a discrete grid. Here, it represents the energy transfer term (nearest neighbor interaction) of the Hamiltonian $H_\varepsilon$.*

## B. Spectral approach in 2D

Here we outline the basic logic of the spectral approach. Detailed proofs and physical interpretation of the model can be found in [41]–[43]. Note that in the general introduction of spectral theory, $v_0$ and $v_1$ are any two (different) vectors in the Hilbert space of interest. Later on, in Sec. III, we chose $v_0 = \delta_0$ and $v_1 = \delta_1$, where $\delta_0$ and $\delta_1$ are the standard basis vectors introduced in equation (1).

The energy spectrum for a given lattice system can be obtained by diagonalizing the corresponding Hamiltonian operator. The 2D discrete random Schrödinger operator in (1) can be diagonalized with the help of the Spectral Theorem, which provides a spectral decomposition of the Hilbert space on which the Hamiltonian acts. It can be shown that cyclicity of all vectors in $l^2(\mathbb{Z}^2)$ is related to the singular part of the spectrum (i.e. localized states) and that non-cyclicity of some vector corresponds to the existence of an absolutely continuous part of the spectrum (i.e. extended states) [44].

<u>Cyclicity</u>: A Hamiltonian $H_\varepsilon$ on a Hilbert space $\mathcal{H}$ is said to have a *cyclic* vector $v_0$ if the span of the vectors $\{v_0, H_\varepsilon v_0, (H_\varepsilon)^2 v_0, \ldots\}$ is dense in $\mathcal{H}$.

<u>Spectral Theorem</u>: When $H_\varepsilon$ is self-adjoint and cyclic, a unitary operator $U$ exists so that

$$H_\varepsilon = U^{-1} M_\xi U, \tag{4}$$

where $M_\xi f(\xi) = \xi f(\xi)$ is the multiplication by the independent variable on another square-integrable Hilbert space $L^2(\mu)$. The new space $L^2(\mu)$ stands for the square-integrable functions with respect to $\mu$ and allows decomposition of the spectral measure $\mu$ into an absolutely continuous part and a singular part

$$\mu = \mu_{ac} + \mu_{\sin g}. \tag{5}$$

The space $L^2(\mu)$ itself is decomposed into two orthogonal Hilbert spaces $L^2(\mu_{ac})$ and $L^2(\mu_{\sin g})$. The Hamiltonian has a part $(H_\varepsilon)_{ac}$ that comes from $L^2(\mu_{ac})$ and a part $(H_\varepsilon)_{\sin g}$ that corresponds to $L^2(\mu_{\sin g})$.

<u>Theorem</u> [45]: For any vector $v_0$ in the lattice space, $v_0$ is cyclic for the singular part $(H_\varepsilon)_{\sin g}$ with probability 1.

<u>Theorem</u> [41]: If one shows that $v_0$ is *not* cyclic for $H_\varepsilon$ with non-zero probability, then almost surely[2]

---

[2] *Note that in probability theory an event happens <u>almost surely</u> if it happens <u>with probability</u> 1. In this paper, we use the two phrases interchangeably.*

$$(H_\varepsilon)_{\sin g} \neq H_\varepsilon \tag{6}$$

which indicates the existence of extended states.

Spectral method: For a given realization of the disorder $W$ in the system:

(i) Fix a random vector, say $v_0$ in the 2D space and generate the sequence $\{v_0, H_\varepsilon v_0, (H_\varepsilon)^2 v_0, \ldots, (H_\varepsilon)^n v_0\}$ where $n \in \{0,1,2,\ldots\}$ is the number of iterations of $H_\varepsilon$ and is used as a timestep.

(ii) Apply a Gram-Schmidt orthogonalization process (without normalization) to the members of the sequence and denote the new sequence $\{m_0, m_1, m_2, \ldots, m_n\}$.

(iii) Calculate the distance from *another* vector in the lattice space, say $v_1$, to the $n$ dimensional orthogonal subspace $\{m_0, m_1, m_2, \ldots, m_n\}$, given by

$$D^n_{\varepsilon,W} = \sqrt{1 - \sum_{k=0}^{n} \frac{\langle m_k | v_1 \rangle^2}{\|m_k\|_2^2}}, \tag{7}$$

where $\|\cdot\|_2$ is the Euclidean norm and $\langle \cdot | \cdot \rangle$ is the inner product in the space. It can be shown [41], [42] that extended states exist with probability 1 if

$$\lim_{n \to \infty} D^n_{\varepsilon,W} > 0 \tag{8}$$

can be shown to be true with a nonzero probability.

## III. RESULTS

### A. Numerical simulation

We applied the spectral approach to the cases of 2D honeycomb, triangular, and square lattices (Fig. 1) using the discrete random Schrödinger operator $H_\varepsilon$ given in equation (1). Of course, $\mathbb{Z}^2$ is now replaced by the honeycomb / triangular lattice. Each simulation was run for $n = 4500$ iterations of the Hamiltonian. The difference in geometry was reflected in the Laplacian, which has the form $\Delta = -ZV$, where $Z$ is the number of nearest neighbors. The hopping potential $V$ was taken to be unity. Thus, the results from the numerical simulations are in units of $V$. For each lattice geometry, we considered disorders $W = 0.10:0.05:1.20$. To facilitate the discussion, in this work, we divide this range of values into small ($W < 0.60$), medium ($0.60 \leq W \leq 0.90$), and large ($W > 0.90$) disorder. This convention is specific for our particular study and will be used throughout the paper.

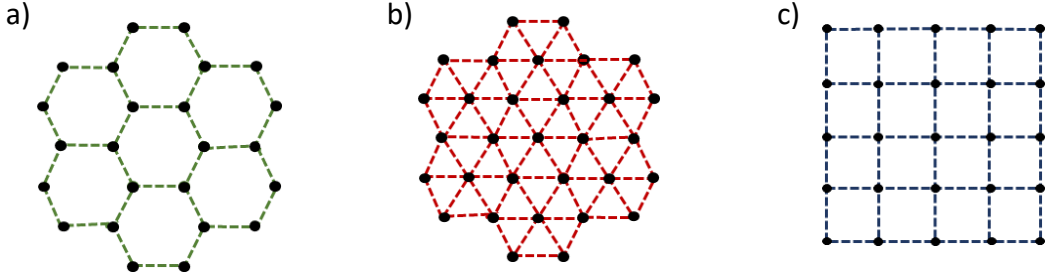

FIG. 1. (Color online) Honeycomb (a), triangular (b), and square (c) lattice geometry. Notice that the triangular symmetry is obtained by placing a lattice point at the center of each hexagon in the honeycomb lattice.

Once the dimension, geometry, and amount of disorder are specified, the simulation generates one realization of each random variable $\varepsilon_i \in [-W/2, W/2]$, which yields the operator $H_\varepsilon$. Next, a vector $\delta_0$ in the 2D space is fixed and used to generate the sequence $\{\delta_0, H_\varepsilon \delta_0, (H_\varepsilon)^2 \delta_0, ..., (H_\varepsilon)^n \delta_0\}$, where $n = 4500$. The simulation then applies the Gram-Schmidt orthogonalization process (without normalization) to the sequence and calculates a value for $D$ after each iteration of the Hamiltonian using equation (7) with $v_1 = \delta_1$. Here one iteration of the Hamiltonian corresponds to the time needed for the energy of the particle in a given state to propagate to its nearest neighbors; thus, the plot of $D$ versus $n$ is analogous to determining the evolution of the distance $D$ over time. Figure 2 shows an (intuitive) visual representation of this process for the triangular lattice.

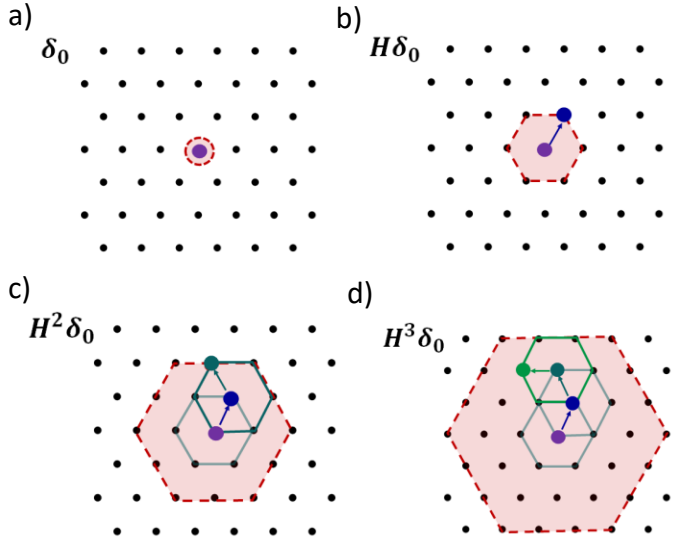

FIG. 2. (Color online) Visual representation of the first three iterations of the Hamiltonian on a triangular lattice. In each picture, the shaded area indicates the number of particles which have potentially been affected by the corresponding iteration of the Hamiltonian. The enlarged dots connected with arrows and the small hexagons show one possible path for propagation.

The particle in (a) is in some initial energy state. After one iteration of the Hamiltonian (b), energy has been transferred to its nearest neighbors. In the second iteration (c), the nearest neighbors (the vertices of the shaded hexagon in (b)) can also transfer energy to their nearest neighbors, including back to the original particle. A subsequent iteration is shown in (d).

To minimize numerical error and ensure randomness of the assigned values of $\varepsilon_i$, five realizations were generated for each $W$ and the resulting distances averaged at each iteration. Figure 3 shows the distance-time evolution of the (averaged) values of $D$ obtained for each geometry. A qualitative examination of the plots indicates that for all three lattices the distance values for small disorder ($W < 0.60$) quickly flatten out and appear as horizontal lines at $n = 4500$. The corresponding log-log plots for these disorders (see Fig. 4) are straight lines, suggesting that the distances are exponentially[3] decreasing towards finite nonzero values. According to the spectral approach, such behavior corresponds to the existence of extended states with probability 1.

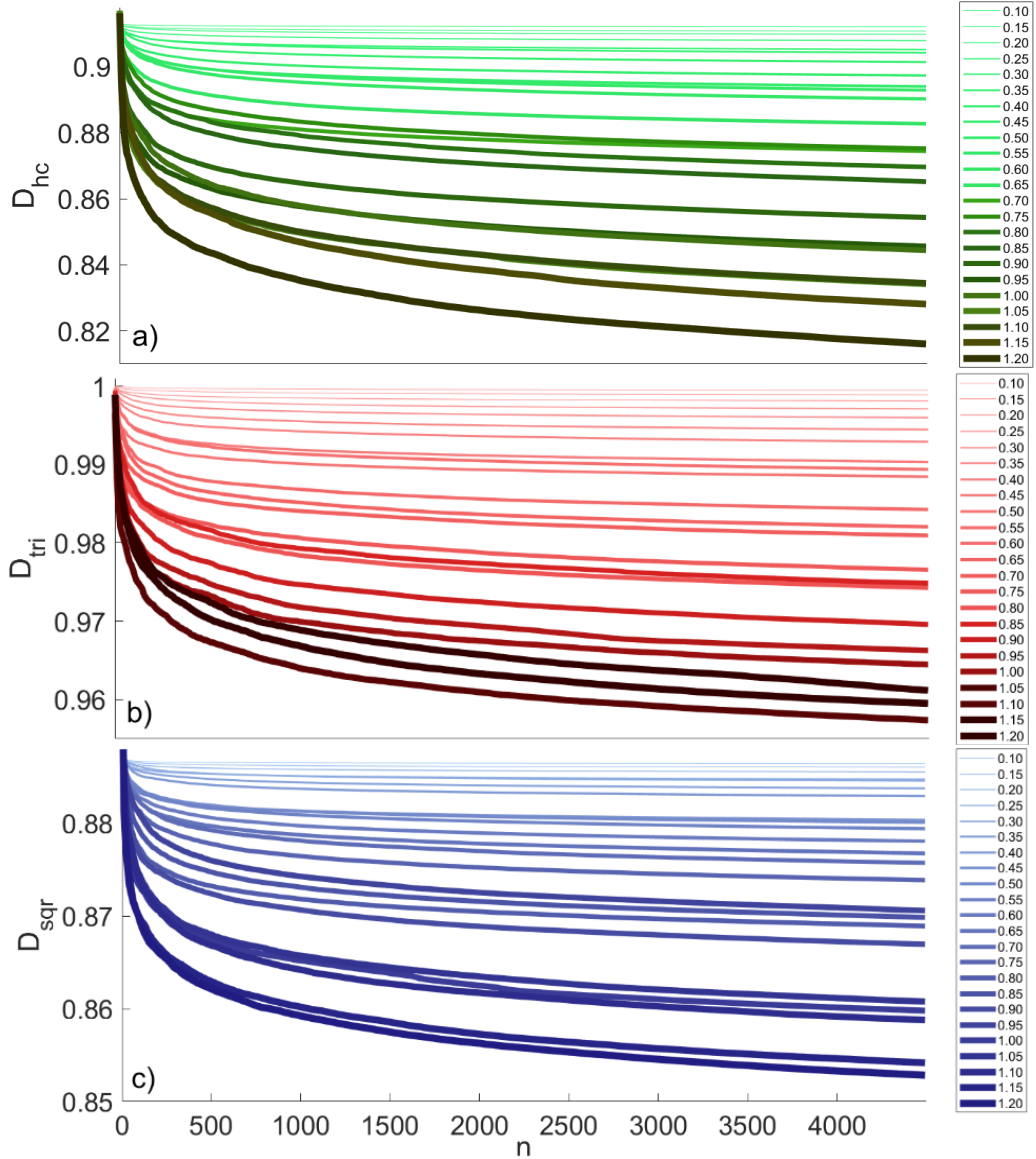

FIG. 3. (Color online) Distance time evolution plots for the honeycomb a), triangular b), and square c) lattices with disorder $W = 0.10 : 0.05 : 1.20$ $W = 0.10 : 0.05 : 1.20$. The data shown is averaged over five different trials.

As the amount of disorder is increased, the distance plots exhibit greater negative slopes and a quantitative analysis is needed to determine the transition from delocalized to localized states. In this analysis, the transition point represents the critical value of disorder for which the spectral approach can no longer show the existence of extended states with probability 1. We emphasize that this transition point does not necessarily indicate a sharp MIT; rather, it corresponds to the critical amount of disorder that marks the onset of the phase transition in the system. It has been previously argued by Thouless and Last [47] that (in both 2D and 3D) there is a transition interval of disorder values $[W_{min}, W_{max}]$ (rather than a single point) for which the behavior of the wavefunction is characteristic of neither extended states nor exponential decay. Here we identify an analog of the lower bound $W_{min}$ for such an interval. The upper bound is the critical amount of disorder above which the wavefunction decays exponentially and is the quantity commonly studied in the literature [48]–[50]. The intermediate region $W_{min} < W < W_{max}$, where the wavefunction may decay following a power law or logarithmically, is of significance for 2D materials, such as graphene.

---

[3] *Note that in the spectral approach an exponential decay of the distance plots to a finite nonzero value corresponds to delocalization. This <u>should not</u> be confused with the exponential decay of the electron wavefunction in scaling theory, which corresponds to localization. The D value is a mathematical construct that is used to test for cyclicity, not a physical wavefunction.*

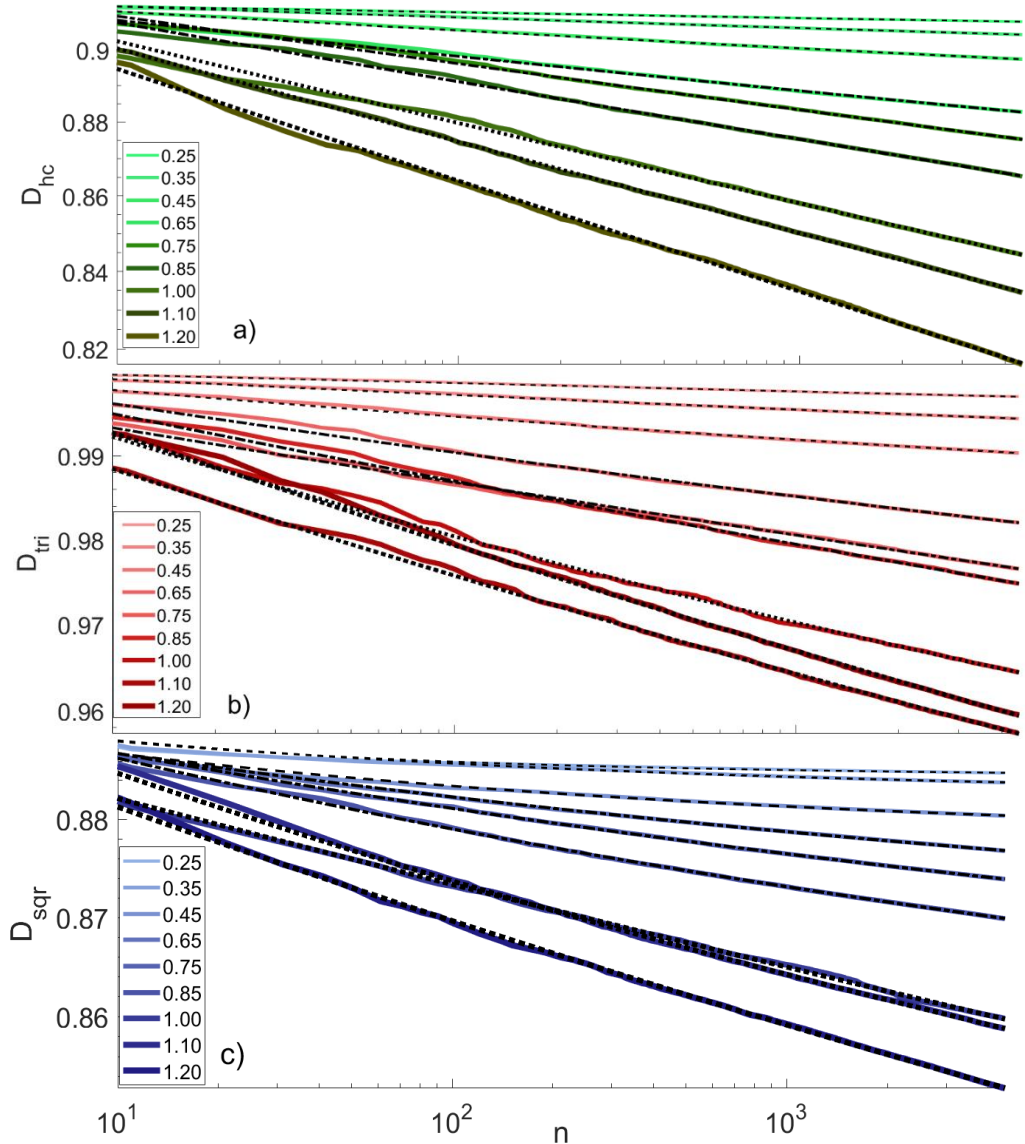

FIG. 4. (Color online) Log-log plots for the honeycomb a), triangular b), and square c) lattices. For clarity of presentation, each graph displays only characteristic trends of the log-log plots for small disorder $W = 0.25, 0.35, 0.45$ ), medium disorder (W = $0.65, 0.75, 0.85$), and large disorder ($W = 1.00, 1.10, 1.20$). In each plot, the dashed, dash-dotted, and dotted lines represent the fit to a nonlinear regression model (see Sec. III C).

## B. Error estimates

As mentioned in the previous section, we generated and averaged five different realizations for each examined disorder value. In each set of five realizations, the standard deviation from the mean value was used as error estimate in the quantitative analysis of the data. Figure 5 shows distance plots of the average values of $D_{hc}$ (solid lines) together with the corresponding error estimates (shaded regions) for three disorder values for the honeycomb geometry. Small error estimates correspond to a small spread of $D$ and higher certainty in the limiting distance value, whereas increasing error estimates indicate significant fluctuations in $D$ and less certainty in the limiting value. As expected, the spread of the random realizations increases with increasing disorder, which indicates an onset of a transition in the limiting behavior of the distance values.

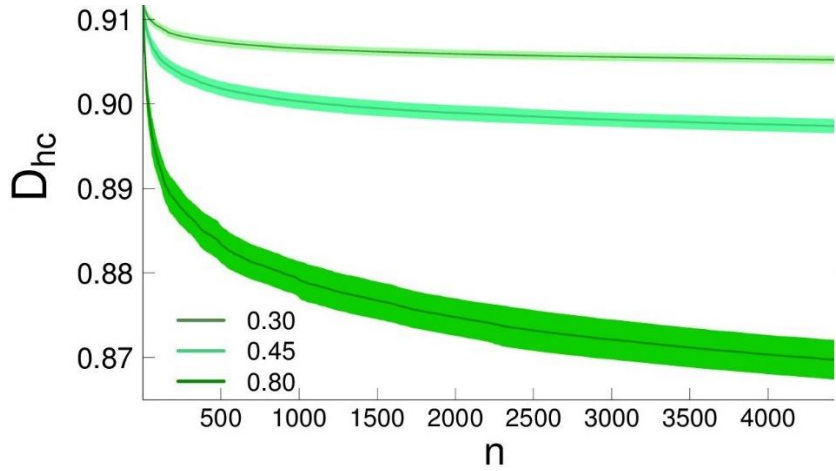

FIG. 5. Error estimates for small, medium, and large disorder values in the graphene lattice. The line gives the average distance at each iteration, while the shaded area indicates the standard deviation of the

## C. Equation fitting

For a given disorder, extended states exist if the corresponding distance parameter $D$ approaches a nonzero value at infinity. As we need to extrapolate the data for $n \to \infty$, we fit the data using an equation of the form

$$D = mn^{-\alpha} + b \tag{9}$$

where the exponent term indicates how rapidly $D$ tends to a finite value and $b$ corresponds to the limiting value of $D$ as $n \to \infty$. We applied a nonlinear regression model to equation (9) and performed hierarchical clustering of the resulting values for $b$, which allowed us to distinguish the delocalized regime for each geometry.

### 1. Nonlinear regression model

Figure 3 shows that the values of $D$. are rapidly changing for $n < 1000$; accordingly, the data fit was performed using a nonlinear regression model for equation (9) with a weight function handle $w = 1/\sqrt{4501 - n}$. To further reduce data fluctuations, separate fits were generated for each of the five realizations of each considered disorder. The resulting values for $b$ were then averaged and the standard deviation from the mean was used as an error estimate. Representative fits for small

($W < 0.60$), medium ($0.60 \leq W \leq 0.90$), and large ($W > 0.90$) disorder are shown in Fig. 4. The extrapolated values of $b$ as $n \to \infty$ are given in Table I. For all considered cases the root mean squared error[4] from the fit equation was consistently small ($\sim 10^{-6} - 10^{-5}$), which indicates a good agreement with the weighted regression model.

Examination of the column containing the values of $b$ for the honeycomb lattice indicates that these values experience a sharp drop when the disorder increases from $W = 0.70$ to $W = 0.75$ (highlighted in Table I). This suggests the existence of two regions of disorder, where the distance values have distinct limiting behavior. The first region corresponds to the delocalized regime and $W = 0.75$ marks the onset of the phase transition to a different transport behavior. Such sharp drops in the distance values are not as obvious for the triangular and the square lattice. However, regions corresponding to distinct behavior of the $b$ values can be identified for all three geometries using a hierarchical clustering analysis introduced in the following section.

TABLE I. Equation parameters yielding the best fit for various amounts of disorder in the 2D honeycomb, triangular, and square lattices. Here $R = \left(D(4500) - b\right)/D(4500)$ measures the relative contribution of the exponential term at $n = 4500$.

| W | Honeycomb $b\,(\times 10^{-3})$ | $R$ (%) | Triangular $b\,(\times 10^{-3})$ | $R$ (%) | Square $b\,(\times 10^{-3})$ | $R$ (%) |
|---|---|---|---|---|---|---|
| 0.10 | $900 \pm 2$ | $1 \pm 0$ | $986 \pm 3$ | $1 \pm 0$ | $886 \pm 0.1$ | 0 |
| 0.15 | $897 \pm 8$ | $1 \pm 1$ | $984 \pm 8$ | $1 \pm 1$ | $886 \pm 0.3$ | 0 |
| 0.20 | $885 \pm 10$ | $3 \pm 1$ | $986 \pm 10$ | $1 \pm 1$ | $885 \pm 0.1$ | 0 |
| 0.25 | $877 \pm 15$ | $3 \pm 2$ | $978 \pm 10$ | $2 \pm 1$ | $884 \pm 1$ | 0 |
| 0.30 | $871 \pm 21$ | $4 \pm 2$ | $978 \pm 13$ | $2 \pm 1$ | $884 \pm 0.5$ | 0 |
| 0.35 | $839 \pm 35$ | $7 \pm 4$ | $967 \pm 21$ | $3 \pm 2$ | $882 \pm 0.5$ | 0 |
| 0.40 | $850 \pm 34$ | $6 \pm 4$ | $957 \pm 15$ | $4 \pm 2$ | $881 \pm 0.6$ | 0 |
| 0.45 | $825 \pm 56$ | $8 \pm 6$ | $968 \pm 13$ | $2 \pm 1$ | $878 \pm 1$ | 0 |
| 0.50 | $833 \pm 42$ | $7 \pm 5$ | $957 \pm 29$ | $3 \pm 3$ | $876 \pm 2$ | 0 |
| 0.55 | $756 \pm 63$ | $15 \pm 7$ | $933 \pm 32$ | $6 \pm 3$ | $873 \pm 1$ | $1 \pm 0$ |
| 0.60 | $700 \pm 55$ | $21 \pm 6$ | $952 \pm 22$ | $3 \pm 2$ | $866 \pm 7$ | $1 \pm 1$ |
| 0.65 | $745 \pm 49$ | $16 \pm 6$ | $922 \pm 56$ | $6 \pm 6$ | $864 \pm 4$ | $1 \pm 1$ |
| 0.70 | $756 \pm 87$ | $14 \pm 10$ | $908 \pm 84$ | $7 \pm 9$ | $844 \pm 30$ | $4 \pm 3$ |
| 0.75 | $585 \pm 76$ | $33 \pm 9$ | $850 \pm 58$ | $13 \pm 6$ | $854 \pm 4$ | $2 \pm 1$ |
| 0.80 | $483 \pm 57$ | $44 \pm 7$ | $830 \pm 35$ | $15 \pm 4$ | $850 \pm 7$ | $2 \pm 1$ |
| 0.85 | $517 \pm 72$ | $40 \pm 8$ | $867 \pm 76$ | $11 \pm 8$ | $842 \pm 15$ | $3 \pm 2$ |
| 0.90 | $641 \pm 110$ | $25 \pm 13$ | $839 \pm 84$ | $13 \pm 9$ | $825 \pm 21$ | $5 \pm 2$ |
| 0.95 | $486 \pm 149$ | $42 \pm 18$ | $831 \pm 162$ | $14 \pm 17$ | $833 \pm 6$ | $4 \pm 1$ |
| 1.00 | $246 \pm 67$ | $71 \pm 8$ | $765 \pm 87$ | $21 \pm 9$ | $771 \pm 81$ | $11 \pm 9$ |
| 1.05 | $265 \pm 198$ | $68 \pm 24$ | $803 \pm 130$ | $16 \pm 14$ | $780 \pm 63$ | $9 \pm 7$ |
| 1.10 | $344 \pm 239$ | $59 \pm 29$ | $828 \pm 68$ | $14 \pm 7$ | $806 \pm 4$ | $6 \pm 1$ |
| 1.15 | $173 \pm 107$ | $91 \pm 18$ | $735 \pm 85$ | $23 \pm 9$ | $763 \pm 53$ | $11 \pm 6$ |
| 1.20 | $155 \pm 179$ | $81 \pm 22$ | $801 \pm 102$ | $16 \pm 10$ | $737 \pm 67$ | $13 \pm 8$ |

The second value in Table I is the ratio $R = \left(D(4500) - b\right)/D(4500)$, which quantifies the contribution of the exponential term in equation (9) to the value of $D$ at $n = 4500$. Although the spectral model identifies the existence of extended states for *any* nonzero limiting value of $D$ (i.e. the exact magnitude of $D$ as $n \to \infty$ is irrelevant), the simulations are finite ($n = 4500$) and the value of $R$ gives a good idea of how rapidly the distance $D$ approaches the limiting value. Small $R$ indicates rapid decay of $D$ to its limiting value $b$, whereas increasing $R$ suggests that the contribution of the exponential term is still significant after 4500 iterations. In the range of disorders for which $R$ is large, the number of iterations may not be sufficient to extrapolate the limiting behavior of $D$ at infinity. For both the honeycomb and the triangular lattice, $R$ increases with increasing disorder and we can again identify the emergence of two regions (corresponding to smaller $R$ and larger $R$), where the rate of decay of $D$ is different. In the case of the square lattice, the ratio remains small for almost all values of disorder considered, suggesting that the transition point for this geometry will be apparent if higher disorder values are included. In the next section, we identify the regions of distinct behavior of $R$ using hierarchical clustering.

---

[4] Note that there are two distinct errors in the discussion. The <u>error estimates</u> obtained from the spread of the random realizations for each disorder (the ones shown in Table I) indicate the certainty with which we can determine the limiting behavior of the distance values. The <u>root mean squared error</u> shows the goodness of the fit.

## *2. Hierarchical clustering*

Since every finite numerical simulation has limitations, a nonzero positive value for $b$ is not the only criterion used in our analysis. Here, we claim the existence of extended states if, in addition to $\lim_{n\to\infty} b > 0$, the following two trends in the distance plots are observed: (i) consistency in the $b$ values, and (ii) consistency in the $R$ values. For each geometry, we identify the region where extended states exist using hierarchical clustering of the values of $b$ and $R$ together with the corresponding error estimates (from Table I). The clustering algorithm uses a Euclidean metric and Ward's minimum variance method. The results for each geometry are represented by the dendrograms in Fig. 6. Dendrograms can be interpreted in two distinct ways: in terms of large-scale groups and in terms of variation among individual branches. The plots in Fig. 6 show the existence of two large-scale clusters for both $b$ and $R$ in each geometry case, which indicates that all examined 2D lattices experience a transition from one transport regime to another as disorder increases. The left cluster in each dendrogram for $b$ (Fig. 6a, b, c) groups together limiting distance values that exhibit small variation with increasing disorder and therefore correspond to the regime where extended states exist. The right clusters mark the formation of a second group of $b$ values which exhibit distinctly different behavior from the first one. Data points within the second cluster correspond to values of disorder which trigger the onset of a phase transition towards different transport behavior. The dendrograms for $R$ (Fig. 6c, d, e) confirm the trends established for the $b$ values.

Comparison between Fig. 6a and Fig. 6b indicates key similarities between the triangular and the hexagonal lattices. In both cases, the left cluster includes all points in the range 1-13 (corresponding to $0.10 \leq W \leq 0.70$) and the first point included in the right cluster is point 14 ($W = 0.75$). In addition, each main cluster of both lattices has two sub-clusters, which group together similar disorder values. Thus, we conclude that in the honeycomb and triangular cases, extended states exist for $W \leq 0.75$. In contrast, the transition in the square lattice begins with point 19 ($W = 1.00$), i.e. extended states in this geometry exist for $W \leq 0.95$ [5]. However, since the right cluster in Fig. 6c consists of only four points, more data should be generated to confirm the transition point of the square lattice.

---

[5] *In our previous work [43], we showed delocalization for $W \leq 0.60$ in the square lattice. This result was obtained with a less robust method for data fit. The weaker method was compensated for by including a worst-case analysis argument. Altogether, this resulted in less resolution for the value of critical disorder. Here, we have improved on this result by generating more data and refining the fitting criteria.*

It is interesting to note that for the honeycomb lattice there is an obvious dissimilarity between the two transport regimes (represented by the difference in cluster heights), which suggests an abrupt phase transition. In contrast, for the triangular case the heights of the two clusters are similar and for the square case, the right cluster is slightly lower than the left one. This indicates a more gradual transition in those two geometries. Such trends in the dendrograms suggest that the "sharpness" of the transition between transport regimes is affected by the number of nearest neighbors, which varies in each geometry.

A limitation of the current analysis is the number of realizations generated and averaged for each disorder value. Based on our previous work, we expect five realizations to be sufficient to identify the global regions of distinct transport behavior, i.e. to distinguish between localized and extended states. However, it is possible that occasionally, the randomly generated five realizations may not be 'representative' of the true behavior of the corresponding distance value. From both Table I and Fig. 6 we see that in the square lattice, $W = 1.10$ (point 21) fall in the left cluster even though it is expected to appear in the right one. We assume that these 'outliers' result from the small number of realizations and do not affect significantly the group behavior of the clusters. Notice that the spectral method was inconclusive for those values since they are considered to lie past the transition points for that geometry.

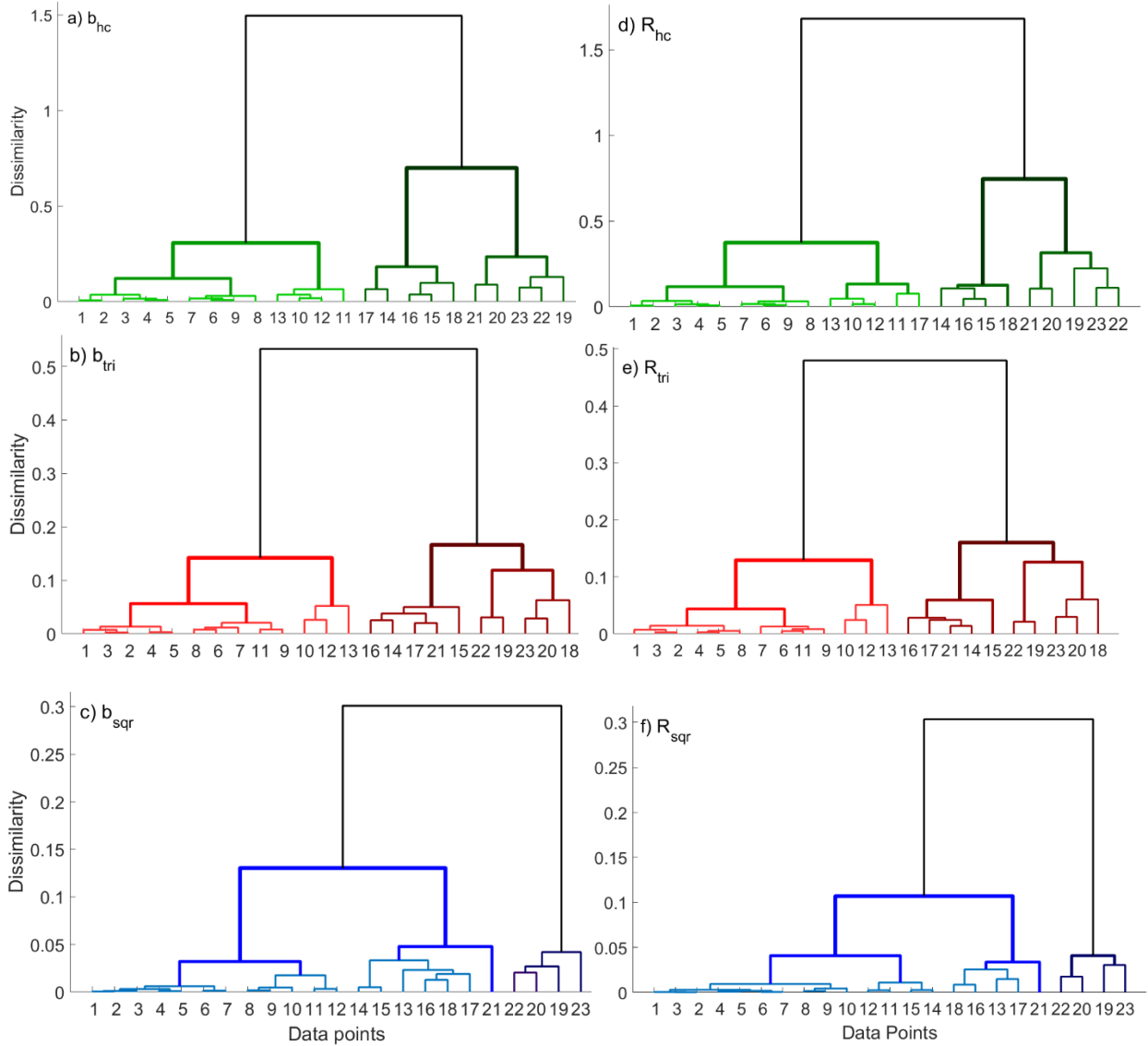

FIG. 6. (Color online) Dendrograms for the $b$ and $R$ values in the honeycomb, triangular, and square lattices. Each ordinal number on the horizontal axis corresponds to a given level of disorder, i.e. $1 = 0.10, 2 = 0.15, 3 = 0.20$, etc. The height of each branch point (or clade) on the vertical axis represents the dissimilarity between clusters connected by that point. The dissimilarity criterion in Ward's method is the total within-cluster error sum of squares, which increases as we move up the tree.

IV COMPARISON BETWEEN THE HONEYCOMB AND THE TRIANGULAR LATTICE

In Sec. III C we demonstrated that the honeycomb and triangular lattices experience onset of a phase transition for similar amounts of disorder. This phenomenon arises due to planar duality between the two geometries. Let $\Lambda$ be a planar honeycomb graph in 2D space. The planar dual $\Lambda^*$ of the honeycomb lattice is the graph constructed by placing a vertex at the center of every face of $\Lambda$. Connecting only pairs of vertices corresponding to adjacent faces shows that the dual of the honeycomb lattice is the triangular lattice (Fig. 7a).

In the Anderson localization problem, one can reflect the amount of disorder by either varying the on-site energies while keeping the hopping integral constant, or by fixing the on-site energies and allowing the hopping potential to vary. In a topological representation of a lattice, the lattice sites correspond to vertices and the hopping integrals – to bonds (or edges). Since every edge of the honeycomb lattice is crossed by a unique edge of the triangular lattice, there is a one-to-one correspondence between the bonds of the two. Thus, if an edge in the real lattice is open (closed), then the dual edge that crosses it, can also be defined as open (closed). It is reasonable to expect that the resulting transport problem in dual space will produce similar results, up to a transformation. The precise transition points and local behavior of the honeycomb and the triangular lattice differ due to the different number of nearest neighbors. However, the behavior of the two lattices in the extended states regime should be highly similar based on the following argument.

Graphically, an extended state can be represented by a cluster of open paths connecting the origin to infinity, while a localized state corresponds to a finite cluster of open paths or a loop. Figure 7b shows a finite cluster of open paths (solid green arrows) starting from the origin $O$. The cluster is finite because it is surrounded by closed edges (light solid green lines), i.e. transitions that are not allowed. Each closed edge is crossed by a unique edge of the dual lattice (red dashed lines). It has been proven [51] that if a finite open cluster contains the origin, then the corresponding edges of the dual form a closed loop, which also contain the origin. Conversely, if a closed loop in the dual contains the origin, then the corresponding open cluster (containing the origin) is finite. Thus, the existence of an infinite path from the origin to infinity in the real lattice occurs when the closed loop of the dual lattice stretches to infinity. From this discussion, it is reasonable to conclude that the delocalization properties of a honeycomb lattice can be examined with the help of a system with triangular symmetry and vice versa.

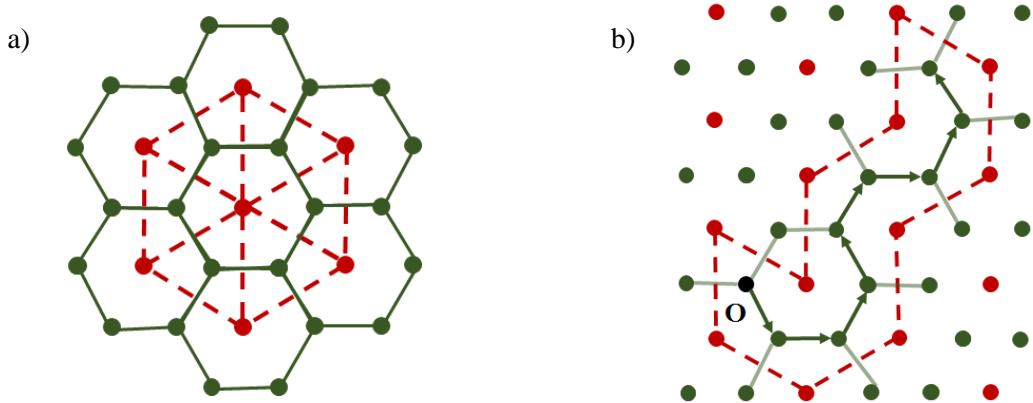

FIG. 7. (Color online) (a) The triangular lattice can be constructed as a dual of the honeycomb lattice. (b) If a finite open path in the honeycomb lattice (indicated by arrows) contains the origin $O$, then there exists a closed loop of closed edges (dashed lines) that also contains the origin. Here the lighter lines between points represent closed edges where transitions are not allowed.

## V. CONCLUSIONS AND FUTURE WORK

In this paper, we applied the spectral approach to delocalization in the 2D honeycomb, triangular, and square lattices. Our numerical simulations established that, in contrast to the predictions of scaling theory, extended states exist for nonzero disorder in all three lattices. The spectral approach, combined with nonlinear regression fitting and hierarchical clustering analysis indicates that delocalization occurs for $W \leq 0.75$ in the honeycomb and triangular lattices, and $W \leq 0.95$ in the square lattice. These results confirm that the existence of metal-to-insulator transition in 2D is independent of the three types of geometries considered here. We also showed that the spectral model correctly predicts the similarity in the transport properties of the honeycomb and the triangular lattices, which is to be expected from planar duality. This justifies the application of the spectral model to systems with triangular symmetries, which are commonly used as analogues to honeycomb lattices, such as graphene. Finally, we observed that the abruptness of the transition from one transport regime to another is dependent on the lattice geometry.

The main goal of this work is to identify the delocalization regime for the three 2D lattices, which was obtained by examining the two main clusters in each dendrogram in Fig. 6. However, the analysis can be extended by considering the substructure of the two main clusters in each geometry. Smaller clusters allow us to identify sub-regimes within the global transport behavior. In our future work, we will include more data points and consider greater values of the disorder, which will improve the accuracy of the statistical analysis and allow us to recognize a rich substructure of transport properties. Since the improved statistical analysis applicable to both classical and a quantum system, below we propose two directions of future research.

### A. Complex plasma crystal as an analogue for graphene

As discussed in Sec. I, graphene is a truly 2D material with extraordinary properties and numerous applications in technology and industry. However, until the transport properties of graphene are well understood, the material cannot be easily utilized as a semiconductor. One approach to resolving this issue is the use of 2D toy systems, or analogues, exhibiting similar properties to graphene. These realizations of 'artificial graphene' are usually hexagonal or hexagonal-like lattices designed to study the tunneling of electrons, atoms, and waves under controlled system conditions [52]. Germane to this argument, the electron-electron interaction in graphene has been shown to be very weak [53], allowing its transport properties to be studied employing classical crystals of similar geometry. One such system, which is easily realized experimentally, is a 2D complex plasma crystal, where micron-sized particles form a triangular lattice in a weakly ionized plasma [54]–[57]. Both electromagnetic and acoustic waves can be induced in 2D complex plasma crystals, which makes them an ideal candidate for the study of Anderson localization. In addition, dusty plasma systems allow control over the fundamental parameters of each scattering center as well as system disorder, providing rapid characterization of the criteria necessary for localization across large parameter spaces. A goal of our future research is to expand the spectral method into the classical regime, which will allow for the use of dusty plasma crystals as a tool for the investigation of transport properties in real materials.

### B. Application to quantum percolation

It has been shown [58], [59] that the Anderson model belongs to the same universality class as the so called quantum percolation problem, which describes a quantum particle moving through a disordered system. The transport problem outlined in Sec. II A can be related to an independent site percolation problem if we consider the 2D integer lattice, where each lattice site is called a *vertex* and two vertices at a Euclidean distance one unit apart are called *neighbors*. Two neighbors are connected by a *bond*. In a *site percolation*[6] problem all bonds are considered open, whereas the vertices are, independently of each other, chosen to be open with probability $p$ and closed with probability $1-p$. An *open cluster* is a set of open vertices. In the case of an infinite size lattice, one is interested in the probability that there exists an open cluster $C(0)$ from the origin to infinity, i.e. the probability that the system percolates. The *percolation probability* (or percolation function) $\theta(p)$. has limiting values $\theta(p=0)=0$ (all vertices closed) and $\theta(p=1)=1$ (all vertices open). Therefore, there exists a critical occupation probability $p_c$ at which the system exhibits a phase transition.

---

[6] *Alternatively, we can consider all vertices to be open and let the bonds be open or closed with a certain probability. This setup is called a <u>bond percolation</u> problem.*

There is a disagreement on the value of the critical probability in 2D, which is related to the 2D Anderson localization problem in the following way. Quantum percolation deals with the problem of a quantum particle moving through a disordered system. The interference of different phases accumulated by the particle as it moves along different paths can lead to Anderson localization. If all states are localized in an infinite 2D crystal for any nonzero disorder (as predicted by scaling theory), then the particle can percolate (delocalize) only when there is zero disorder in the crystal corresponding to critical probability $p_q = 1$ (all sites open, all bonds open). However, the existence of extended states for a nonzero disorder (established by the spectral approach) indicates that the quantum particle will percolate for $p_q < 1$. Thus, resolution of the 2D Anderson localization problem will contribute greatly to the study of the exact value of the critical probability in quantum percolation theory.

The spectral approach to delocalization indicates that for small amounts of the disorder (i.e., $W$ smaller than some critical value), extended states exist almost surely. The relationship between the percolation probability and the amount of disorder, $p(W)$, is dependent on the type of disorder and the choice of distribution function, $\chi(\tau)$, that assigns the disorder to the lattice sites. In this paper, we used a constant distribution for $\chi(\varepsilon)$ and varied only the amount of disorder in the system. Similarly, one can fix the amount of disorder and vary the choice of a distribution function, or both $W$ and $\chi(\varepsilon)$ can be varied. In future work, we will compare the square, bimodal, and Gaussian distributions for various amounts of disorder in 2D systems to study the dependence of the percolation probability on the amount of disorder for each distribution.

## ACKNOWLEDGEMENTS

This work was supported by the Simons Foundation (grant number 426258, Constanze Liaw) and NSF/DOE (NSF grant numbers 1414523, 1740203, 1262031, Lorin S Matthews and Truell W Hyde).

All authors greatly acknowledge the contribution of Dr. Amanda Hering, who gave us valuable advice on the statistical analysis of our data.